\documentclass[aps,twocolumn,superscriptaddress,showpacs]{revtex4}
\usepackage{dcolumn}
\usepackage{bm}
\usepackage{graphicx}
\usepackage{color}
\DeclareGraphicsExtensions{. jpg,. pdf, . mps, . png, . eps, . ps, . EPS}
\begin{document}
\def\be{\begin{equation}}
\def\ee{\end{equation}}

\def\bc{\begin{center}} 
\def\ec{\end{center}}
\def\bea{\begin{eqnarray}}
\def\eea{\end{eqnarray}}
\newcommand{\avg}[1]{\langle{#1}\rangle}
\newcommand{\Avg}[1]{\left\langle{#1}\right\rangle}
 \newcommand{\CHR}[1]{{\tt [#1 - CHR]}}

\title{A unified framework for quasi-species evolution and stochastic quantization }
\author{ Ginestra Bianconi} 

\affiliation{Department of Physics, Northeastern University, Boston, 
Massachusetts 02115 USA}
\author{Christoph Rahmede}
\affiliation{Department of Physics \& Astronomy, 
University of Sussex, Brighton BN1 9QH, UK}
\begin{abstract}
In this paper we provide a unified framework for quasi-species evolution  and stochastic quantization. We map the biological  evolution  described by the quasi-species equation to the  stochastic dynamics of an ensemble of particles undergoing a creation-annihilation process.
We show that this mapping identifies a natural decomposition of the probability that an individual has a certain genotype into eigenfunctions of the evolutionary operator. This alternative approach to study the quasi-species equation allows for a generalization of the Fisher-theorem equivalent to the Price equation. According to this relation the average fitness of an asexual population changes with time proportional to the variance of the eigenvalues of the evolutionary operator.
 Moreover,  from the present alternative formulation of stochastic quantization a novel scenario emerges to be compared with existing approaches.
 The evolution of an ensemble of particles undergoing diffusion and a creation-annihilation process  is parametrized by a variable $\beta$ that we call the inverse temperature of the stochastic dynamics.
We  find that the evolution equation at high  temperatures is simply related to the Schr\"odinger equation but at low temperature it strongly deviates from it.  In  the presence of additional noise in scattering  processes  between  the particles, the evolution reaches a steady state described by the Bose-Einstein statistics. 
\end{abstract}
\pacs{89.75.-k, 87.23.Kg,05.10.Gg}


\maketitle
\section{Introduction}
The intriguing relation between evolutionary dynamics and statistical mechanics \cite{Fisher,Kimura,Hirsh,Gerland} has attracted the interest of classical evolutionary theorists. Fisher \cite{Fisher,Price} and Kimura \cite{Kimura} have related their results to the second principle of thermodynamics and to the theory of gases. Interestingly, the relation between evolutionary theory \cite{Nowak} and quantum statistical mechanics is emerging  from  a series of independent works \cite{Kingman,Krug,Bose,Fermi,Complex,W,Kadanoff,Ferretti,Shraiman} that show a class of  phase transitions occurring in the evolution of haploid populations and other evolving complex systems described  by a Bose-Einstein condensation. In haploid populations, this transition is the quasi-species phase transition \cite{Nowak,Eigen,Sigmund,Gillespie, Hartl,Sequence} in which a finite fraction of an asexual population ends up having the same genotype if the selective pressure is higher than a  critical value and the mutation rate is smaller than a critical value.
Moreover, in a recent paper \cite{Bose_n} it has been shown that a condensation transition in the Bose-Einstein universality class occurs also in the evolution of diploid sexual populations in presence of epistatic interactions. When this condensation occurs, a finite fraction of pairs of genetic loci in epistatic interactions is fixed.
 
The deep relation between evolutionary theory and quantum mechanics formalisms extends also to the dynamical description of  biological  evolution. The quantum spin-chain formalism \cite{Baake}  and the Sch\"odinger equation in imaginary time \cite{Ebeling_book,Hanggi1,Hanggi2} have been shown to  solve models of asexual evolution. The role of path integrals in describing the temporal evolution of populations \cite{Peliti,Leibler,Lassig} has been  highlighted recently.

Biological evolution is essentially a stochastic process. The relation between quantum mechanics and stochastic dynamics has been deeply explored over the  years \cite{Risken}. In particular the  existing theory of stochastic processes for diffusing particles is described by Langevin and Fokker-Planck equations. Stochastic quantization approaches  define stochastic diffusing processes whose probability density converges to path integrals of some specific quantum system \cite{SQ,Jona}. Stochastic quantization has attracted large interest in the physics community and has been shown to allow for large numerical simulations of quantum systems \cite{sqbook}.
In this paper we propose a unified framework for quasi-species biological evolution and  stochastic quantization.

In the first part of the paper we describe the evolution of an asexual population with overlapping generations in the limit of large populations. We show that it is possible to describe the evolution of a population in terms of eigenfunctions of the evolutionary dynamics, and this description allows for a generalization of the Fisher theorem in presence of mutations. We show that the variation in time of the average fitness of the population is proportional to the variance of the eigenvalues of the evolutionary dynamics. Therefore the average fitness of the population  asymptotically in time reaches the highest eigenvalue of the evolutionary dynamics. This relation is very general and applies as well in absence of mutations, for low mutation rate and for high mutation rate. Therefore it is not in general related to the localization  of the eigenfunction  with maximal fitness on the fitness landscape. Moreover, in the first part of the paper related to biological evolution we review the results of the Kingman model which are related to the emergence of quantum statistics in the equation of biological evolution.
These results are related to the quasi-species description of evolutionary dynamics that neglects genetic drift, which has significant relevance in presence of small populations and small mutation rate $\mu N\ll 1$ or in the cases in which the population is fluctuating and undergoes population bottlenecks.

In the second part of the paper we  consider an ensemble of elementary particles which do not only diffuse, like in existing approaches of stochastic quantization, but undergo also creation and annihilation processes mimicking the dynamics of biological evolution. The probability density $P(x,t)$ that a generic particle is at position $x$ at time $t$ follows the quasi-species equation \cite{Nowak} first proposed by Eigen \cite{Eigen} to describe biological evolution. This allows us to define a new dictionary for a mapping from the evolutionary theory to the stochastic dynamics of particles undergoing the creation-annihilation process. The ensemble of particles is subjected to a stochastic Gaussian noise and a potential energy. The stochastic noise can be mapped to the mutations of biological evolution and the potential energy can be mapped to a Fisher fitness function.  Moreover the stochastic evolution strongly depends on the parameter $\beta$, that we call the {\em inverse temperature} of the system, playing the equivalent role of {\em selection pressure} in the biological evolutionary mapping. The probability of a certain configuration of the ensemble of particles satisfies the quasi-species equation and can be decomposed into eigenfunctions of the evolutionary operator with a discrete spectrum. For low values of $\beta$, $\beta \ll 1$ (i.e., low {\em selection pressure}) the quasi-species equation is directly related to the Schr\"odinger equation, while for $\beta\gg1$ (i.e., high {\em selection pressure}) the quasi-species equation strongly deviates from the Schr\"odinger equation. In order to solve in one case the quasi-species equation for all possible values of the inverse temperature we explicitly solve the case of an harmonic potential. The evolution turns out to be dissipative, with a relaxation of the configuration to the fundamental state. 
Moreover we investigate the role of genetic drift in this model, showing that thanks to the gaussian nature of the noise the systems always reaches a stationary state. In particular, asymptotically in time  the spatial distribution of particles fluctuates around the  fundamental state predicted by the solution of the quasi-species equation. The smaller is the number of particles the stronger are the fluctuations aournd the mean.
Moreover we include an additional noise describing a scattering process of the particles such that after each scattering process a particle takes a random position distributed according to a fixed probability distribution. When we introduce this scattering probability the ensemble of particles is proven to follow a Bose-Einstein distribution. Interestingly, the relaxation occurring in absence of the scattering process obeys a dynamical equation which has its equivalent in the generalization of the Fisher theorem found in the mean-field treatment of the biological evolution presented in the first part of this paper. In this mapping  the ground state of the ensemble of particles corresponds to the state of maximal reproductive rate of the asexual population. Therefore the relaxation of the ensemble of particles to  the ground state corresponds to the relaxation  to the state corresponding to the maximal eigenvalue of the  evolution of an asexual population. Finally, in presence of the scattering process the Bose-Einstein distribution is emerging naturally with similarities to the Kingman model \cite{Kingman} for the evolution
  of asexual populations with infinite number of genetic loci.

\section{Evolution of asexual populations}
\subsection{The quasi-species equation}
The genome of an asexual organism is formed by a single copy of each chromosome.
If we indicate by $i=1,\ldots , N $ a genetic locus, a given genotype is determined by
the allelic states $\{\sigma\}=(\sigma_1,\sigma_2,\ldots,\sigma_i,\ldots \sigma_N)$ at each genetic locus $i$.
 The allelic state $\sigma_i$ at each genetic locus  $i$ can take 4 values corresponding to adenine, thymine, cytosine, guanine, i.e.  $ \sigma_i=1,2,3,4$.
The  population evolves under the drive of selection that favors allelic configurations corresponding to higher reproduction rate,  and mutations that increase the genetic variation in the population.
We assume that the reproductive rate $W(\{\sigma\})$ also called Wright fitness of a genotype $\{\sigma\}$ is given by 
\begin{equation}W(\{\sigma\})=e^{-\beta U(\{\sigma\})}\end{equation} where $U(\{\sigma\})$ is the Fisher fitness and $\beta$ is the {\em selective pressure}. If $\beta=0$ every genotype has the same reproductive rate. If $\beta\gg1$ the difference in the reproductive rate of  genotypes having different $U(\{\sigma\})$ is strongly enhanced.

We assume that at each time there can be a birth or a death process. We assume that the birth process depends on the reproductive rate $W(\{\sigma\})=\exp[-\beta U(\{\sigma\})]$ and that the death process is a random drift.
If we define the probability $P(\{\sigma\},t)$ that at time $t$ an individual has genome $\{\sigma\}$,
the dynamical equation of evolution of $P(\{\sigma\},t)$ is  given by the mean-field equation, valid for large populations or growing populations, i.e. 
\be
\frac{d P(\{\sigma\},t)}{dt}={\mathbf M}_{\{\sigma\}|\{\sigma^{\prime}\}}\left[\frac{e^{-\beta U(\{\sigma^{\prime}\})}P(\{\sigma^{\prime}\},t)}{Z_t}\right]-P(\{\sigma\},t)
 \label{ev}
\ee
where  the operator ${\mathbf M}_{\{\sigma\}|\{\sigma^{\prime}\}}$ is defined in $(\ref{M1})$ and $Q(\{\sigma\}|\{\sigma^{\prime}\})$ is given by 
\begin{equation}
Q(\{\sigma\}|\{\sigma^{\prime}\})=\prod_i\left[(1-\mu)\delta(\sigma_i, \sigma^{\prime}_i)+\frac{\mu}{4}\right]
\label{Q1}
\end{equation}
where $\mu$ is the mutation rate.
 The partition function $Z_t$ in Eq. $(\ref{ev})$ is given by
\begin{equation}
Z_t=\sum_{\{\sigma\}}\sum_{\{\sigma^{\prime}\}}Q({\{\sigma\}|\{\sigma^{\prime}\}})e^{-\beta U(\{\sigma^{\prime}\})}P(\{\sigma^{\prime}\},t).
\label{defZ00}
\end{equation}
We observe here that the  equation $(\ref{ev})$ can be reduced  to the well-known quasi-species  equation  \cite{Nowak,Sigmund,Gillespie, Hartl,Sequence} if we make a change of variables $t\to t^{\prime}$ with $dt/Z_t=dt^{\prime}$.
Let us now assume to know the solution of the eigenvalue problem 
\begin{equation}
{\mathbf M}_{\{\sigma\}|\{\sigma^{\prime}\}}\left[e^{-\beta U(\{\sigma^{\prime}\})}\pi_n(\{\sigma^{\prime}\})\right]=\lambda_n \pi_n(\{\sigma\})
\label{eigp1}
\end{equation}
with $\lambda_n$ describing the discrete spectrum of this problem and $\pi_n(\{x\})$ the normalized eigenfunctions.
If we decompose the function $P(\{\sigma\},t )$ on the basis $\pi_n(\{\sigma\})$ of eigenfunctions , i.e.
\begin{equation}
P(\{\sigma\},t)=\sum_n c_n(t) \pi_n(\{\sigma\}),
\end{equation}
and if we make the self-consistent assumption that we know the function $Z_t$,
the dynamical solution of Eq. $(\ref{ev})$  for the coefficients $c_n(t)$  is given by 
\begin{equation}
c_n(t)=\exp\left[ {\lambda_n}G(t)-t\right]c_n(0).
\label{dyn2}
\end{equation}
In Eq. $(\ref{dyn2})$ the function $G(t)$ is defined through the function $Z_t$ according to the equation
\begin{eqnarray}
G(t)=\int_0^t dt^{\prime}\frac{1}{Z_{t^{\prime}}} \ .
\label{G}
\end{eqnarray}
Using the definition for $Z_t$ given by Eq. $(\ref{defZ00})$ together with Eqs. $(\ref{Q1})$  and $(\ref{dyn2})$, we can  close the self-consistent equations and  uniquely  determine the evolutionary dynamics of the population. Therefore the partition function $Z_t$ is given by
\begin{eqnarray}
Z_t=\sum_n \lambda^n c_n(t)=\avg{\lambda}.
\end{eqnarray}
where the average $\avg{\cdot}$ is performed over the functions $c_n(t)$ given by $(\ref{dyn2})$.
Finally, using $(\ref{dyn2})$ we can derive the equation obeyed by the  partition function $Z_t=\avg{\lambda}$,
\begin{equation}
\frac{1}{2}\frac{d\avg{\lambda}^2}{dt}=\avg{\lambda^2}-\avg{\lambda}^2.
\label{FD}
\end{equation}
This equation generalizes   the Fisher theorem of natural selection   \cite{Fisher,Leibler} in presence of mutations and describes the fact that evolution is an off-equilibrium process.
In fact, $\avg{\lambda}=\Avg{W(\{\sigma\})}$ is the average reproductive rate of the population. Eq. $(\ref{FD})$ expresses the fact that this average reproductive rate, asymptotically in time,  evolves toward the fundamental state $\lambda_0$, as long as the environment does not change, i.e.  the Fisher fitness function $U(\{\sigma\})$ and the selective pressure remain constant in time. 
The rate at which the average fitness change is equal to the variance of the eigenvalues $\lambda_n$ over the distribution $c_n(t)$.
In the Fisher theorem, valid in absence of mutations, the average fitness increases proportionally to the variance of the individual fitnesses.
Therefore we see that the generalization of the Fisher theorem \cite{Fisher}, equivalent to the  Price equation \cite{Price}, can be done substituting the variance of the individual fitnesses with the variance of the eigenvalues of the evolutionary dynamics. We note here that since $c_n(t)$ sum up to one, but they are not necessarily positive definite, in presence of mutations, the fitness of the population can either increase or decrease.
In any case, asymptotically in time the population is described by the eigenfunction associated to the maximal eigenvalue $\lambda_0$.
We observe here that the Price equation corresponds to the decomposition of the Eq. $(\ref{FD})$ in a term independent of the mutation rate and a term 
dependent on the mutation rate.
Moreover we observe that the Fisher theorem is not only valid in presence of constant fitness landscape, considered in this paper, but is also valid in presence of fluctuating environments. Indeed  most   of the recent literature on the Fisher theorem relates to fluctuating environments, where $U(\{\sigma\})$ is not constant. For  recent results in these interesting aspects of biological evolution  we refer the reader to the recent papers \cite{Leibler, Lassig}.

\subsection{ The Bose-Einstein condensation in the  Kingman model }

The Kingman model   \cite{Kingman,Krug} is one of the most interesting stylized models of asexual evolution where the quasi-species transition is observed.
In the framework of this model the quasi-species transition can be exactly mapped to the Bose-Einstein condensation in a Bose gas.
One interesting aspect of the Kingman model is that the evolutionary dynamics reaches an equilibrium due to the constant drive of random mutations. 
In the Kingman model each individual is assigned a single real  parameter $\epsilon\geq 0$ determining its reproductive rate $W(\epsilon)$, i.e. 
\begin{equation}
W(\epsilon)=e^{-\beta \epsilon}.
\end{equation}
Moreover, in this model, after each duplication, a   mutation occurs with probability $\alpha$ and a new offspring is generated  with random fitness $\epsilon$ drawn  from a given distribution $\rho(\epsilon)$.

Therefore instead of writing Eq. $(\ref{ev})$ for the distribution $P(\{\sigma\},t)$ that an individual of the population has genotype $\{\sigma\}$ at time $t$, we can write the equation for the probability density ${ P}(\epsilon,t |t_0)$ that a random individual in the population is associated with a  given Fisher fitness  $\epsilon$ at time $t$ and had the last mutation at time $t_0$.
The evolution of $P(\epsilon,t |t_0)$ is given by 
\be
\frac{d P(\epsilon,t|t_0)}{dt}=(1-\alpha)\frac{e^{-\beta \epsilon}P(\epsilon,t|t_0)}{Z_{t}}-P(\epsilon,t|t_0).
 \label{evB}
\ee
The partition function $Z_{t}$ in Eq. (\ref{evB}) is given by 
\be
Z_{t}=\int_0^t dt_0 \int d\epsilon\, e^{-\beta \epsilon}P(\epsilon,t|t_0).
\label{defZB}
\ee
Asymptotically in time we assume self-consistently that $Z_{t}\rightarrow Z$.
Therefore, in this limit  the   probability $P(\epsilon,t|t_0)$ that an individual has  fitness $\epsilon$ at time $t$ under the condition that the last  mutation happened at time $t_0$, is given by the solution of $(\ref{evB})$, i.e.
\begin{equation}
P(\epsilon,t|t_0)=\alpha \rho(\epsilon) e^{\left[\exp(-\beta \epsilon) (1-\alpha)\frac{1}{Z}-1\right](t-t_0)}\ .
\end{equation}
Finally, integrating over $t_0$, we can evaluate the  probability $P(\epsilon,t)$ that an individual  at time $t$ has a given Fisher fitness  $\epsilon$ independently of $t_0$, i.e.
\begin{equation}
P(\epsilon,t)=\int_0^{t} dt_0\alpha \rho(\epsilon) e^{\left[\exp(-\beta \epsilon) (1-\alpha)\frac{1}{Z}-1\right](t-t_0)}.\nonumber
\end{equation}

Therefore the  
steady state solution for ${P}^{\star}_B (\epsilon)$  reached in the limit $t\to\infty$   is given by 
\begin{equation}
{P}^{\star}_B(\epsilon)=\alpha\rho(\epsilon)\left[1+\frac{1}{{e^{\beta (\epsilon-\mu_B)}}-1}\right]
\label{BE}
\end{equation}
where $e^{-\beta \mu_B}=Z/(1-\alpha)$
and the  probability that an individual has fitness $\epsilon$ is determined by the Bose-Einstein distribution.
Finally, using the definition Eq. $(\ref{defZB})$ we can find the self-consistent equation that the constant $Z=(1-\alpha)e^{-\beta \mu_B}$ needs to satisfy, i.e.
 the  normalization condition
\begin{equation}
\frac{1-\alpha}{\alpha}=\int d\epsilon \rho(\epsilon)\frac{1}{{e^{\beta (\epsilon-\mu_B)}}-1}.
\label{BE2}
\end{equation}
If $\rho(\epsilon)$ vanishes for $\epsilon\to 0$, the Bose-Einstein integral in Eq. $(\ref{BE2})$ can be limited from above. As a result, at  high enough selection pressure and low enough mutation rate the system might undergo a condensation phase transition in the Bose-Einstein universality class. 
Below this phase transition a finite fraction of the individuals in the population shares the same  genotype corresponding to  the maximal fitness.
This is one of the principal examples that show the so called quasi-species transition: for low mutation rate and high selection a finite fraction of the population is found to have the same genotype. Below this phase transition the system is not stationary anymore and the population average fitness is strongly dependent on the statistic of records related to the occurrence of  largest fitness.
Interestingly a similar phase transition occurs also in evolving ecologies   \cite{Ferretti},  where the invasive species might strongly reduce the biodiversity, and in  evolving models of  complex networks   \cite{Bose,W}, where there might be the emergence of super-hubs like Google in the World-Wide-Web.

\subsection{The Fermi-Dirac  distribution in presence of negative selection}
The Fermi-Dirac distribution   can be obtained in presence of negative selection  by a similar mechanism that generates Bose-Einstein distribution in the Kingman model.
This model is mostly interesting because of the underlying symmetry between Fermi-Dirac and Bose-Einstein distributions.
We assume that each individual is assigned a parameter  $\epsilon$ describing its adaptability to the environment. The birth rate is one and new individuals are generated at each  unit of time  $t$ with parameter $\epsilon$ drawn from a given distribution $\rho(\epsilon)$. The  death rate   is given by a random drift with probability $\alpha$ and by a negative selection proportional to $\exp[\beta \epsilon]$ with probability $1-\alpha$.  
Therefore  the dynamical evolution of the probability density $P(\epsilon,t|t_0)$ that in the population there is an individual with parameter  $\epsilon$ born at time $t_0$ is given by 
\be
\frac{d P(\epsilon,t|t_0)}{dt}=-\left[(1-\alpha) \frac{e^{\beta \epsilon}}{Z_{t}}+\alpha\right]P(\epsilon,t|t_0)
 \label{evF}
\ee
The partition function $Z_{t}$ in Eq. (\ref{evF}) is given by 
\be
Z_{t}=\int_0^t dt_0 \int d\epsilon e^{\beta \epsilon}P(\epsilon,t|t_0)\ .
\label{defZF}
\ee
Assuming in the asymptotic limit  of large times   $Z_{t}\rightarrow Z$ to solve Eq. $(\ref{evF})$ we find that the probability density that an individual has fitness $\epsilon$ at time $t$ under the condition that it was born at time  $t_0$ is  given by 
\begin{equation}
P(\epsilon,t|t_0)= \rho(\epsilon) e^{\left[\exp(\beta \epsilon) (1-\alpha)\frac{1}{Z}-\alpha\right] (t-t_0)} .
\end{equation}
The probability density that an individual at time $t$ has Fisher fitness   $\epsilon$ independently of $t_0$ is given by 
\begin{equation}
P(\epsilon,t)=\int_0^t dt_0 \rho(\epsilon) e^{\left[\exp(-\beta \epsilon) (1-\alpha)\frac{1}{Z}-\alpha \right](t-t_0)}.\nonumber
\end{equation}

This systems reaches the steady state in the limit $t\to \infty $ with the equilibrium distribution given by the Fermi-Dirac distribution
  ${P}_F^{\star} (\epsilon)$, and in particular we have 
\begin{equation}
{ P}^{\star}_F(\epsilon)=\frac{1}{\alpha}\rho(\epsilon)\frac{1}{e^{\beta (\epsilon-\mu_F)}+1}
\end{equation}
with  $e^{\beta \mu_F}=Z\alpha/(1-\alpha)$.
This distribution is the Fermi-Dirac distribution.
The self consistent argument is closed by determining self-consistently the  equations that $Z$ has to satisfy, i.e. the normalization condition
\begin{equation}
\alpha=\int d\epsilon \rho(\epsilon)\frac{1}{e^{\beta (\epsilon-\mu_F)}+1}.
\end{equation}
The duality between Fermi-Dirac and Bose-Einstein distribution has been also recognized in the framework of evolving network models   \cite{Complex, Fermi} and of models for evolving  ecologies   \cite{Ferretti}.

\section{ Evolution of an ensemble of particles } 
\subsection{Evolution of a large ensemble of particles described by the quasi-species equation}
We now study an ensemble of particles located in a one-dimensional space  that undergoes a creation-annihilation process. 
We map the evolution of the particles to biological evolution. In this mapping, the fitness function corresponds to  the energy of a particle, and mutations correspond to  a stochastic noise. 
Finally  we assume that  the probability  to find a particle at a certain space-time point for large ensemble of particles obeys the quasi-species equation.
We simply replace in Eq. (\ref{ev}) the biological population by a large ensemble of particles and the individual genome by a point in continuous one-dimensional space. 

In the following, we  define $P(x,t)$ as the probability density that a particle is at position $x$ at time $t$. We assume that this distribution, in the limit of large number of particles,  follows the  stochastic equation inspired by Eq. (\ref{ev}) for biological evolution
\begin{equation}
\frac{d P(x,t)}{dt}=\frac{{\mathbf M}_{x,x^{\prime}}\left[e^{-\beta U(x^{\prime})}P(x^{\prime},t)\right]}{Z_t}-P(x,t)
 \label{ev3}
\end{equation}
where the partition function $Z_t$ is given by
\be
Z_t=\int dx^{\prime}\int dx \, Q(x,x^{\prime}) e^{-\beta U (x^{\prime}))}P(x^{\prime})
\label{defZp}
\ee
and the operator  ${\mathbf M}_{x,x^{\prime}}$, applied to a function $e^{-\beta U(x)}f(x)$, acts as
\begin{equation}
{\mathbf M}_{x,x'}\left[e^{-\beta U(x^{\prime})}f(x^{\prime})\right]=\int dx^{\prime} \, Q(x,x^{\prime})e^{-\beta U(x^{\prime})}f(x^{\prime}) \nonumber
\label{M1}
\end{equation}
where $Q(x,x')$ describes the stochastic noise playing the role of { mutations} for  the evolution of the ensemble of particles.
The inverse temperature $\beta$,    with $\beta>0$, plays in this stochastic dynamics the same role as the selective pressure in the biological evolutionary dynamics.
For simplicity we assume that the position $x$ at time $t+dt$ is related to the position $x^{\prime}$ at time $t$ by $x=x^{\prime}+\eta$ where $\eta$ is a noise with Gaussian distribution. Therefore we take
\begin{equation}
Q(x,x^{\prime})=\sqrt{\frac{1}{2\pi\beta D}}\int d\eta\, \delta(x^{\prime}-x-\eta)e^{-\frac{1}{2 D \beta}\eta^2  }
\label{qr}
\end{equation}
where $D\beta$ is the variance  of the noise. 

Introducing the Fourier representation for the delta function we get
\begin{equation}
\hspace{-4mm}{\mathbf M}_{x,x^{\prime}}\left[e^{-\beta U(x^{\prime})}f(x^{\prime})\right]={\cal N}\int dx^{\prime} \int dk \, e^{-\beta[H(x,k)]+ik (x-x^{\prime})} f(x^{\prime})\nonumber
 \label{S}
\end{equation}
where ${\cal N}=\frac{1}{2\pi}\sqrt{\frac{1}{2\pi \beta D}}$ and the  Hamiltonian $H(x,k)$ of this system is identified as
$H(x,k)=\frac{D}{2 }k^2+U(x)$.
The action of the operator ${\mathbf M}_{x,x^{\prime}}$ on a function $f(x)$
for  $\beta \ll 1$ is given by
\begin{eqnarray}
&&\hspace*{-10mm}{\mathbf M}_{x,x^{\prime}}\left[e^{-\beta U(x^{\prime})}f(x^{\prime})\right]\simeq\nonumber\\
&&\simeq{\cal N}\int dx^{\prime} \int dk \left\{1-\beta H(x,k)\right\}e^{ik (x-x^{\prime})} f(x^{\prime})\ .\nonumber 
 \label{Sa}
\end{eqnarray}
Therefore for $\beta\ll1 $ the evolution Eq. (\ref{ev3}) can be written as 
\begin{eqnarray}
\hspace*{-10mm}\frac{d P(x,t)}{dt}&\simeq&-\frac{\cal N}{Z_t} \beta\left[-\frac{D}{2}\frac{\partial^2 P(x,t)}{\partial x^2 }+
U(x)P(x,t)\right] \nonumber \\
&&+\left(\frac{\cal N}{Z_t} -1\right)P(x,t)\ .
 \label{SA}
\end{eqnarray}
This equation is particularly interesting since it is linear in the Hamiltonian $H(x,k)$ and can be put in relation with the  Schr\"odinger equation in imaginary time. In the next steps of our calculation we consider this particular limit ($\beta\ll1$) to compare the results of our evolutionary dynamics of particles with the  solution of the Schr\"odinger  equation in quantum mechanics.

In order to derive a specific solution of the stochastic dynamics described by Eq. (\ref{ev3})  we consider  the potential energy $U(x)$ of  the harmonic oscillator, i.e.
\begin{equation}
U(x)=\frac{1}{2}m\,\omega^2x^2.
\label{Ux}
\end{equation}
Assuming that time and space dependence of the probability density $P(x,t)$ can be factorized, we solve the following  eigenvalue equation for the space-dependent part of this probability, i.e.
\begin{equation}
{\mathbf M}_{x,x^{\prime}}\left[e^{-\beta U(x^{\prime})} \pi_n(x^{\prime})\right]=\lambda_n \pi_n(x)\ .
\label{eigp}
\end{equation}
The normalized eigenfunctions are given by 
\begin{equation}
\pi_n(x)=B_n(x)e^{-\frac{1}{2D}\epsilon x^2}
\label{eif}
\end{equation}
where   $B_n(x)$ indicates the  polynomials of order $n$ that solve this eigenvalue problem.
Two possible values for $\epsilon=\epsilon_1$ and $\epsilon=\epsilon_2$  solve  this eigenvalue problem, i.e.
\begin{equation}
\epsilon_{1,2}=\frac{\beta D m\omega^2}{2}\left(\pm\sqrt{1+\frac{4}{D\beta^2m\omega^2}}-1\right)\ .
\label{alpha}
\end{equation}
As the square root is always larger than one, one has $\epsilon_1>0$ and $\epsilon_2<0$. The eigenfunctions are only normalizable for positive $\epsilon=\epsilon_1$. 
Then one obtains the eigenvalues
\begin{equation}
\lambda_n=\frac{1}{{(1-\beta\epsilon_{2})^{n+1/2}}} \ .
\label{l}
\end{equation}
Since the eigenvalues $\lambda_n$ are always positive we define the energy spectrum of the system as the set of values $E_n$ given by 
\begin{equation}
\lambda_n=e^{-\beta E_n}.
\label{eiv1}
\end{equation}
Therefore, using Eq. $(\ref{l})$ and Eq.   $(\ref{eiv1})$ we can calculate   the spectrum $E_n$, given by 
\begin{equation}
\beta E_n=\left(n+\frac{1}{2}\right)\log(1-\beta\epsilon_2).
\label{eiv2}
\end{equation}
For $\beta\ll 1$, using Eqs. (\ref{alpha}), we find
\begin{equation}
\epsilon_1=\sqrt{m\omega^2 D}+{\cal O}(\beta)\ ;\qquad
\epsilon_2=-\sqrt{m\omega^2 D}+{\cal O}(\beta)\ .
\end{equation}
Finally,   we can calculate the spectrum $E_n$ of Eq. $(\ref{eiv2})$, at the first order in $\beta$, which is given by
\begin{eqnarray}
E_n=\sqrt{m D}\, \omega\left(n+\frac{1}{2}\right).
\end{eqnarray}
This spectrum  coincides with   the spectrum of the quantum mechanical harmonic oscillator if we put  $D$  proportional to $\hbar^2/m$. 
Let us continue to study the stochastic evolution of every possible value of $\beta$.
If we decompose the function $P(x,t)$ in the basis $\pi_n(x)$, 
\begin{equation}
P(x,t)=\sum_n c_n(t)\, \pi_n(x)\ ,
\label{p}
\end{equation}
we obtain the self-consistent dynamical solution
\begin{equation}
c_n(t)=\exp\left[ {\lambda_n}G(t)-t\right]c_n(0)
\label{dynp}
\end{equation}
where the function $G(t)$ is defined by the function $Z_t$ according to the equation
\begin{eqnarray}
G(t)=\int_0^t dt^{\prime}\frac{1}{Z_{t^{\prime}}} \ .
\label{G2}
\end{eqnarray}
The solution of our evolutionary equation is given by  the definition for $Z_t$ in Eq. $(\ref{defZp})$, the eigenvalues Eq. (\ref{eigp}), and  the  structure of the solution of the probability density $P(x,t)$ given by  Eq. $(\ref{p})$ we  find 
\begin{eqnarray}
Z_t=\sum_n \lambda_n c_n(t)=\avg{\lambda}\ ,
\end{eqnarray}
where  $c_n(t)$ are given by Eq. $(\ref{dynp})$.
Finally using  Eqs. (\ref{dynp}), (\ref{G2}) it is easy to prove that 
\begin{equation}
\frac{1}{2}\frac{d\avg{\lambda}^2}{d t}=(\avg{\lambda^2}-\avg{\lambda}^2) \ .
\end{equation} 
The partition function $Z_t=\avg{\lambda}$  therefore  describes off-equilibrium dynamics.
From the dynamical solution Eq. $(\ref{dynp})$ for the coefficient $c_n(t)$ it is easy to prove that the probability distribution  $P(x,t)$ converges asymptotically in time to the fundamental state $\pi_0(x)$ associated to the largest eigenvalue $\lambda_0$.
This phenomenon has a biological equivalent in Eq. $(\ref{FD})$ according to which a population of asexual individuals has a fitness that reaches the highest eigenvalue of the evolution operator.

We observe that for the ensemble of particles the maximal eigenvalue $\lambda=\lambda_0$  corresponds to the ground state energy $E_0$ according to Eq. (\ref{eiv2}).
Therefore, with time the particle system relaxes to this ground state.

Finally, asymptotically in time, the average energy $\avg{U(x)}$  of the particles is given by the average of $U(x)$ defined in Eq. $(\ref{Ux})$ on the fundamental eigenfunction given by Eq. $(\ref{eif})$. Therefore  we find
\begin{equation}
\Avg{U(x)}=\frac{1}{2}\frac{m\omega^2 D}{\epsilon}.
\end{equation}
In the limit $\beta\ll 1$ we find that the average energy $ \Avg{U(x)}$ is given by 
\begin{equation}
\Avg{U(x)}=\frac{1}{2} \hbar \omega \ .
\end{equation}
In  the limit $\beta\gg 1$ instead, we find that $\Avg{U(x)}$ is given by a classical type of expression, i.e. 
 \begin{equation}
 \Avg{U(x)}=\frac{1}{2}\beta (\hbar \omega)^2
\end{equation}
where we have put $D=\hbar^2/m$.
\subsection{Stochastic effects}

The analytical treatment we have presented in the previous paragraph focus  on the quasi-species equation. It is therefore a mean-field treatment that neglects the stochastic fluctuations present in the dynamics that we are proposing to study.
In biological evolution these effects are known as  {\it genetic drift}.
Typically these effects are important for  small populations $N$ and small mutation rates $\mu$, i.e. $N\mu\ll 1$.
For example in the Kingman model we might have that the fixation time  of the fittest mutant can be smaller than the typical time needed for new mutations to occur. When this occurs   the  evolutionary dynamics strongly depends on the statistics of records at which the fittest mutants arise in the population.

\begin{figure}
\includegraphics[width=80mm, height=60mm]{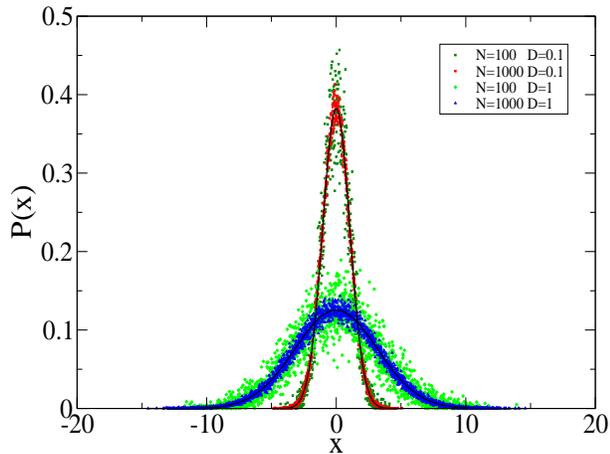}
\caption{(Colour on-line) The asymptotic distribution of an ensemble of $N$  particles  undergoing a Moran process driven by an harmonic potential $U(x)=\frac{1}{2}m\omega^2 x^2$ with $m=1$ and $\omega=1$ and selective pressure $\beta=10$.
The distribution are collected after equilibration of the system for a time window of $400N$ time steps. The solid lines indicate the theoretical expectations based on the quasi-species equation given by Eq. $(\ref{eif})$. The smaller the population the larger are the stochastic fluctuations around the mean behavior described  by the quasi-species equation. }
\label{stochastic}
\end{figure}

In this section we study the stochastic effects in the model of evolution of the ensemble of particles under the influence of an  harmonic potential $U(x)$.
In order to simulate the evolution of particle we adopt the Moran process \cite{Nowak}. We assume to have an ensemble of  $N$ particles. Each particle $i=1,2,\ldots, N$ is assigned a  position $x_i$.
Starting from random initial conditions, at each time step we choose a particle $i$ to generate a new particle.
The particle $i$ is choosen according to the probability
\begin{equation}
\Pi_i=\frac{e^{-\beta U(x_i)}}{\sum_{r=1}^N e^{-\beta U(x_r)}}
\end{equation}
The new particle will have a position $x^{new}$ chosen with probability
\begin{equation}
Q(x^{new},x_i)=\frac{1}{\sqrt{2\pi\beta D}}e^{-\frac{1}{2\beta D}(x_i-x^{new})^2}.
\end{equation}
Finally we choose a random particle $j$ to be annihilated and we put
\begin{equation}
x_j\rightarrow x^{new}.
\end{equation}
In Fig. $\ref{stochastic}$ we show the asymptotic distribution for the ensemble of particles as a function of the noise amplitude $D$ and the 
number of particles $N$ in the ensemble. The fluctuations of these distributions are stronger for small ensemble of particles ($N$ small) and high values of $D$.
Nevertheless always  the quasi-species solution well captures the average of the distribution. In fact  the asymptotic distribution fluctuates around the eigenfunction $\pi_0(x)$ (given by Eq. ($\ref{eif}$)) corresponding to the fundamental states of energy $E_0$.
The suprisingly  good agreement of the quasi-species approach reside  on the fact that this stochastic process is defined over  a continuous variable $x$ and is described by a Gaussian noise. Therefore the diffusion can be arbitrary small but is always present as long as $D\neq 0$.
This is at odds with genetic mutations that might not occur over a finite time frame if the mutation rate is small enough giving rise to a separation of time scales in the system.

\subsection{Evolution in presence of a scattering process } 

In the following we study the change in the evolution of the particles in the presence of a scattering process occurring with probability $\alpha$ where at any time there is a creation process.
In order to mimic this process we assume that when a particle undergoes a scattering process it takes a random  position  distributed according to a probability distribution $g(x)$ which does not depend on time.
In order to study the stochastic evolution of this interacting system of particles we describe the evolution of the ensemble of particles that have not scattered   since time $t_0$.
The evolution for the probability density $P(x,t|t_0)$ that a particle is at position $x$ at time $t$ under the condition that the last scattering happened at time  $t_0$ is given by 
\begin{equation}
\frac{d P(x,t|t_0)}{dt}=(1-\alpha)\frac{{\mathbf M}_{x,x'}e^{-\beta U(x^{\prime})}P(x^{\prime},t|t_0)}{Z_{t}}-P(x,t|t_0).
 \label{evis}
\end{equation}
where ${\mathbf M}_{x,x'}$ is given by Eq. (\ref{M1}) and $Q(x,x^{\prime})$ is given by Eq. (\ref{qr}).
The partition function $Z_t$ in Eq. (\ref{evis}) is given by 
\be
Z_{t}=\int_0^{t} dt_0 \int dx^{\prime}\int dx \, Q(x,x^{\prime}) e^{-\beta U (x^{\prime})}P(x^{\prime},t|t_0)
\label{defZ5}
\ee
with $Q(x,x')$ given by Eq. (\ref{qr}).
In order to solve Eq. $(\ref{evis})$ we have to solve the  eigenvalue Eq. $(\ref{eigp})$ for the interacting case. The eigenvalues and  the eigenfunctions  remain unchanged and are given by Eqs. (\ref{eiv2}), (\ref{eif}).
If we decompose the function $g(x)$ in the basis of eigenfunctions of ${\mathbf M}_{x,x'}$ we get
$g(x)=\sum_n g_n \pi_n(x)\ .$
Asymptotically in time, the dynamics will reach a stationary state determined by a constant limit value $Z_t\rightarrow Z$.
We assume self-consistently to know the value of $Z$.
With this assumption, at large times $t$  we solve the dynamical Eq. $(\ref{evis})$ to find the  probability density $P(x,t|t_0)$ that a particle is at position $x$ at time $t$ under the condition that the last  scattering happened at $t_0$, i.e.
\begin{equation}
P(x,t|t_0)=\sum_n\alpha\, g_n e^{[\lambda_n (1-\alpha)\frac{1}{Z}-1](t-t_0)}\pi_n(x)\ .
\end{equation}
Finally, integrating $P(x,t|t_0)$ over $t_0$ we get the probability $P(x,t)$ that a particle is at time $t$ at position $x$ independently of $t_0$, i.e.
\begin{equation}
P(x,t)=\int_0^{t} dt_0\sum_n\alpha\, g_n e^{[\lambda_n (1-\alpha)\frac{1}{Z}-1](t-t_0)}\pi_n(x)\ .
\label{px}
\end{equation} Using  Eq. (\ref{px}) and neglecting the terms vanishing in the limit $t\to \infty$ we obtain that 
the probability $p(n)$ of  an eigenstate $n$ is expressed in terms of the 
 Bose-Einstein distribution
\begin{equation}
p(n)=g_n\alpha \left[1+\frac{1}{e^{\beta(E_n-\mu_B)}-1}\right]
\end{equation}
which takes the form of a Bose-Einstein distribution with  $e^{-\beta \mu_B}=Z/(1-\alpha)$.
Finally the constant $Z=(1-\alpha)e^{-\beta \mu_B}$ is fixed by the normalization condition
\begin{equation}
1-\alpha=\sum_n g_n \frac{1}{e^{\beta(E_n-\mu_B)}-1}.
\end{equation}
This demonstrates that in this context it is possible  to explain the  emergence of the Bose-Einstein distribution by purely dynamical considerations. In particular this mechanism has its parallel in the Kingman model \cite{Kingman} for the  evolution  of asexual populations with infinite number of genetic loci.

 \section{Conclusions}
 In this paper we have proposed a unified framework to study  biological quasi-species evolution and stochastic quantization.
 We  have shown  that the dynamics of biological evolution of non-overlapping generations and large populations is naturally expressed in terms of eigenvalues and eigenfunctions of the evolutionary dynamics.
 This approach allows for a generalization of the Fisher theorem in presence of mutations. In this framework the average fitness of the population changes in time proportionally to the variance of the eigenvalues instead of the variance of the fitness of the individuals.
 
Moreover we have shown that the  quasi-species equation describing the evolution of a large  ensemble of particles provides  an alternative  path to stochastic quantization with respect to existing approaches. The probability distribution of the ensemble of particles can be decomposed into the eigenfunctions of the evolution operator.  In the low temperature limit the quasi-species equation is related to the Schr\"odinger equation. Therefore,  in this limit the spectrum  coincides with the spectrum of the associated quantum dynamics.
Interestingly, in the presence of a noise mimicking random scattering processes the Bose-Einstein distribution can be reached. 

Direct relations between biological evolution and quantum mechanics were proposed in the book {\em What is life?} \cite{Schroedinger} by Erwin Schr\"odinger. Since then this idea continues to fascinate biologists and physicists and it is gaining momentum in these last years \cite{Davies,Lloyd,Penrose,McFadden}. The relations between quasi-species  evolution and quantum mechanics pointed out here are of formal nature.
 We believe  that this paper  open new perspectives on   the relation between quasi-species evolution, stochastic processes and stochastic quantization that will stimulate further research at the intersection of these fields. 

\section{Acknowlegments}
We acknowledge comments and stimulating discussions with H. Goldberg, G. Jona-Lasinio, M. Kardar, P. Nath M. Nowak and S. Redner.

\end{document}